\documentclass[submission,copyright,creativecommons]{eptcs}
\providecommand{\event}{FMAS 2021} 
\usepackage{graphicx}
\usepackage[most]{tcolorbox}
\usepackage{tabto}

\usepackage{algorithm}  
\usepackage{algorithmic}  
\usepackage{amsmath,amsfonts,amssymb,amsthm}

\usepackage{listings}

\lstnewenvironment{code}[1][]%
    {\hfill \\
    \lstset{basicstyle=\ttfamily\scriptsize,frame=none,#1}{\linewidth}}
    {\endminipage\hfill\null}
    
\definecolor{commentGreen}{RGB}{90,135,100} 
\definecolor{keywordPurple}{RGB}{125,10,65} 

\title{Assuring Increasingly Autonomous Systems in Human-Machine Teams: An Urban Air Mobility Case Study\thanks{We would like to thank Natasha Neogi and Paul Miner of NASA LaRC for their input on Urban Air Mobility scenarios of interest and for their feedback on safety assessment and verification approaches.}}

\author{Siddhartha Bhattacharyya\(^1\), Jennifer Davis\(^2\), Anubhav Gupta\(^3\), Nandith Narayan\(^1\), \\Michael Matessa\(^2\)
\institute{\(^1\)Florida Institute of Technology, Melbourne, FL, USA\\}
\institute{\(^2\)Collins Aerospace, Cedar Rapids, IA, USA\\}
\institute{\(^3\)University of British Columbia, Kelowna, BC, CA\\}
\email{sbhattacharyya@fit.edu, jen.davis@collins.com}
}

\def\titlerunning{Assuring Increasingly Autonomous Systems in Human-Machine Teams: A UAM Case Study}
\def\authorrunning{Bhattacharyya, Davis, Gupta, Narayan, \& Matessa}
\begin{document}
\maketitle

\begin{abstract}
As aircraft systems become increasingly autonomous, the human-machine role allocation changes and opportunities for new failure modes arise. This necessitates an approach to identify the safety requirements for the increasingly autonomous system (IAS) as well as a framework and techniques to verify and validate that an IAS meets its safety requirements. We use Crew Resource Management techniques to identify requirements and behaviors for safe human-machine teaming behaviors. We provide a methodology to verify that an IAS meets its requirements. We apply the methodology to a case study in Urban Air Mobility, which includes two contingency scenarios: unreliable sensor and aborted landing. For this case study, we implement an IAS agent in the Soar language that acts as a copilot for the selected contingency scenarios and performs takeoff and landing preparation, while the pilot maintains final decision authority. We develop a formal human-machine team architecture model in the Architectural Analysis and Design Language (AADL), with operator and IAS requirements formalized in the Assume Guarantee REasoning Environment (AGREE) Annex to AADL. We formally verify safety requirements for the human-machine team given the requirements on the IAS and operator. We develop an automated translator from Soar to the nuXmv model checking language and formally verify that the IAS agent satisfies its requirements using nuXmv. We share the design and requirements errors found in the process as well as our lessons learned.

\end{abstract}

\section{Introduction}
Increasingly, autonomous systems are evaluated to operate with humans for safety, security and mission-critical operations. This is evident from research in multiple domains such as medical, aerospace, and defense. One of the major advantages of using an autonomous agent is the ability to process 
 much more data in real-time than a human can handle. In civil aviation, the level of autonomy of systems is expected to increase gradually over time, hence such systems are referred to as Increasingly Autonomous Systems (IAS) \cite{Neogi2016CapturingSR}. This term is used in the singular form to indicate a system that incorporates more autonomous functions than are in use today. As the level of autonomy increases, the human-machine role allocation changes and there is the opportunity for new failure modes to arise. Therefore, the objective of this work is to develop a framework and techniques for the verification and validation of IAS in novel role allocations. We present the framework and apply it to a case study we developed in Urban Air Mobility.

The vision for Urban Air Mobility (UAM) \cite{uam, Gregory_uam} is to provide flexible, short-distance air travel for the masses. For this to be truly achievable we have to develop increasingly autonomous systems that can handle complex flight operations, including contingency management. Furthermore, to ensure safety of flight, these systems need to be verifiable. Thus, our research effort focuses on the creation of an assurance framework that integrates human-machine interactions with formal-methods-based rigorous analysis, along with simulation. 


 With the increasing complexity and autonomy in systems, traditional verification approaches such as testing face scalability challenges. Our verification and validation 
 approach includes the following tenets: 
\begin{itemize}
\item Use Crew Resource Management to identify requirements and procedures for safe human-machine teaming behaviors
\item Include the human in the model so that human-machine interactions can be analyzed
\item Use formal methods where possible and practical to prove safety requirements are satisfied by (the model of) the system or component
\item Where possible and practical, use automated translation and build tools so that the deployed system is equivalent to the one we analyzed
\item Simulate contingency management scenarios with the target air vehicle to explore potential teaming behaviors and to test the human-autonomy team in conjunction with a high-fidelity model of the vehicle
\end{itemize}



The contributions of this work are the following:
\begin{enumerate}
\item Methodology for the verification and validation of increasingly autonomous systems in human-machine teams

\item Development of a case study in Urban Air Mobility, including:
\begin{enumerate}
\item Realistic UAM example scenarios (unreliable sensor and aborted landing)
\item An IAS agent implemented in Soar that acts as a copilot with increasing role assignment for the selected scenarios as well as takeoff and landing preparation
\end{enumerate}

\item Application of the methodology to the case study, including:
\begin{enumerate}
\item Scenario simulations in X-Plane with a realistic UAM aircraft, the AgustaWestland AW609
\item A formal human-machine team (operator-IAS) architecture model in AADL that supports the two example scenarios. The operator and IAS requirements are formalized in the AGREE Annex to AADL.
\item Formal verification of properties (using AGREE) for the human-machine team given the requirements on the IAS and operator
\item Formal verification of properties (using nuXmv) for the IAS agent 
\end{enumerate}

\item A Soar-to-nuXmv translator\footnote{https://assistresearchlab.fit.edu/}

\end{enumerate}

Our methodology is discussed in Section \ref{sec:background}. Background information on languages, tools, and techniques is provided in Section \ref{sec:prelim}. We describe our UAM case study, including the application of our methodology, results, and lessons learned, in Section \ref{sec:case-study}. Our Soar-to-nuXmv translation algorithm is provided in Section \ref{sec:translator}. Finally, conclusions and future work are discussed in Section \ref{sec:conclusion}.

\section{Methodology}
\label{sec:background}


 One of the fundamental challenges in developing human-level agents is defining the primitive computational structures that store, retrieve, and process knowledge. Equally important is defining the organization of those computational structures. A cognitive architecture provides fixed computational structures that form the architecture of the human mind. It is not a single algorithm or method for solving a problem; it is the task-independent infrastructure that brings an agent's knowledge to solve a problem. 
Cognitive architecture based production systems are a popular method in Artificial Intelligence for producing intelligent behavior that is understandable to the program operator. Common rule-based reasoning systems include the General Problem Solver (GPS) \cite{gps},  the MYCIN knowledge based inference system \cite{mycin}, 
the Adaptive Control of Thought-Rational Theory (ACT-R) \cite{actr} and the Soar cognitive architecture \cite{soar}. 

\begin{figure}[h]
\centering
\includegraphics[scale=0.50]{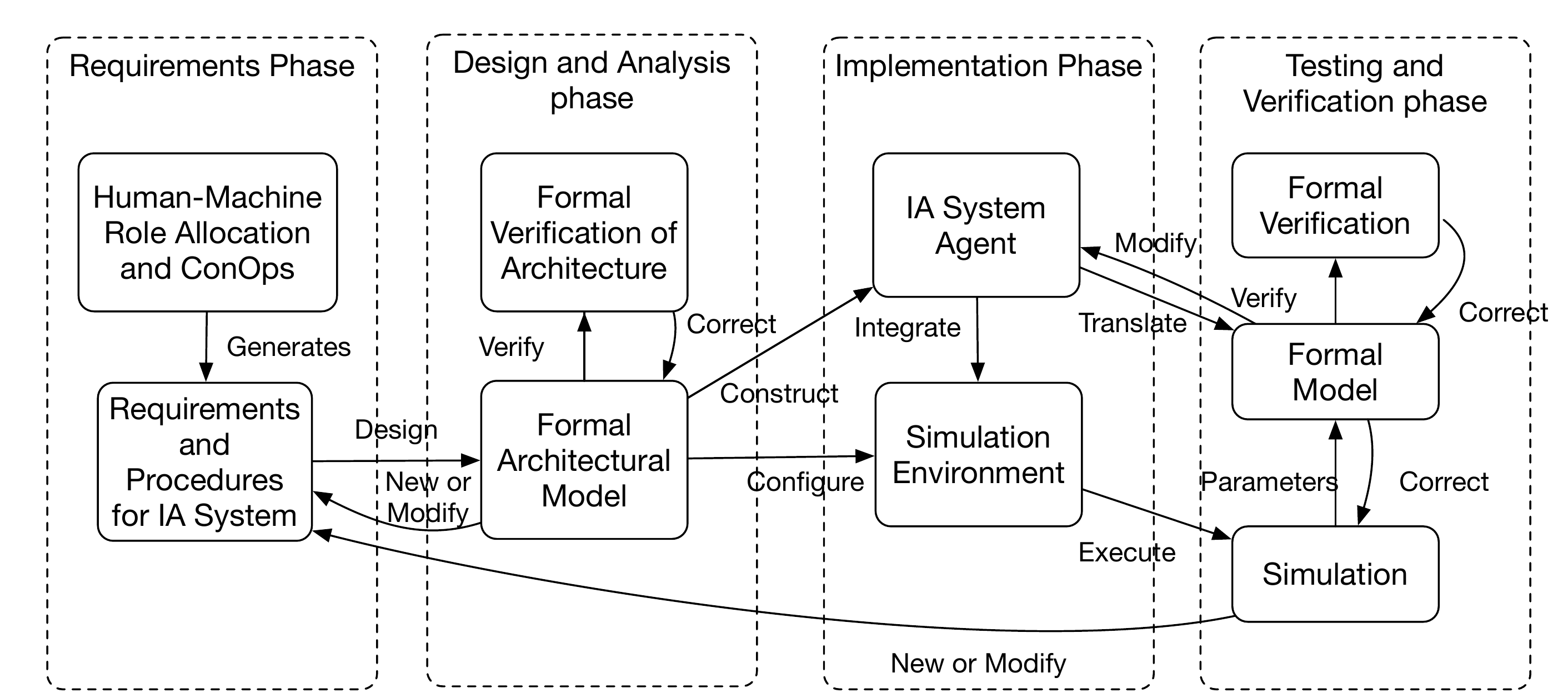}
\caption{AHMIIAS framework}
\label{fig:TechApproach}
\end{figure}

Formal verification of cognitive architecture is a more recent research area, where Langenfeld et al. \cite{Langenfeld2019OnFV} have developed a manual approach to the translation from ACT-R to Uppaal. Previously, Bhattacharyya et al. have developed a framework to automate the translation of Soar to Uppaal \cite{NfmBhattacharyya}. While this was a successfully implemented approach, it lacked the integration of requirements from human-machine interaction research, as well as architectural design and verification, which are both included in this framework.
There are four main steps to formal verification and validation of the human-machine team in our approach. They are identified in Fig. \ref{fig:TechApproach}.
\begin{enumerate}
    \item Requirements phase: Develop scenarios to identify and capture human-machine roles and interactions, and then derive the requirements for the IAS.
    \item Design and analysis phase with formal verification: Create a formal architectural model for the system with human-IAS interactions captured as requirements allocated to the human and IAS components. Perform formal analysis on the architectural model to show that human-IAS team safety properties are satisfied given that the component requirements are satisfied.
    \item Implementation phase: Construct the IAS agent with human-IAS interactions based on the verified architectural model. Also, configure the simulation environment.
    \item Testing and Formal Verification phase: Execute the simulation scenarios generated in the requirements phase to test the satisfaction of requirements by the implementation. Translate the IAS agent behavior to a formal verification environment (nuXmv) to formally verify the behavior. The input parameters generated from the simulation environment are modeled as an input template within the formal verification environment (nuXmv). 
\end{enumerate}

\section{Preliminaries}
\label{sec:prelim}

\subsection{Crew Resource Management}

In current two-pilot commercial operations, collaboration skills are taught as Crew Resource Management (CRM) \cite{flin2017}. The result of this training has been increased safety for the aviation industry. Various researchers have noted the applicability of CRM to Human-Autonomy Teaming \cite{crm, shively2018, taylor2018}. One survey of pilots found the majority agreed that automation should adhere to CRM rules \cite{taylor2017}. For this project, we implement basic CRM skills in the IAS to provide coordinated crew behavior. In the area of communication, we implement the CRM skill of waiting for acknowledgement to ensure that the other crew member has knowledge of the information that was told to them. In the area of management, we implement the CRM role of Pilot In Command who listens to input from other crew members but has the final authority in any decision that is made.

\subsection{Soar}
\label{sec:Soar}

Soar is a general cognitive architecture that provides a computational infrastructure that resembles the cognitive capabilities exhibited by a human. Soar implements knowledge-intensive reasoning that enables execution of rules based on the context. It also has the capability to integrate learning into the intelligent agent using chunking or reinforcement learning. Several rule-based reasoning systems were surveyed as candidates for modeling human-automation interactions \cite{gps,mycin,soar,actr}.  Soar was selected because it encompasses multiple memory constructs (e.g., semantic, episodic, etc.) and learning mechanisms (e.g., reinforcement, chunking etc.).  Soar production rules are expressed in first-order logic, which makes them amenable to verification.  Finally, Soar is a programmable architecture with an embedded theory. This enables executing Soar models on embedded system platforms and studying the design problem through rapid prototyping and simulation.

\subsection{AADL}
\label{sec:aadlintro}
The Architecture Analysis and Design Language (AADL) \cite{feilerAADLbook} is a standardized language designed for embedded, real-time systems. It supports design, analysis, virtual integration, and code generation. It can be used to predict and validate runtime characteristics including security, timeliness, and availability. It comes with an error model annex to support fault modeling and hazard analysis. The Open Source AADL Tool Environment (OSATE) tool developed by SEI provides the modeling environment for developing in AADL. 

\subsection{AGREE}
\label{sec:agree}
One of the barriers to formal verification of large systems is the scalability of the analysis methods and tools. The Assume Guarantee REasoning Environment (AGREE) \cite{AGREE-paper} was developed as a plugin for the OSATE environment to overcome this barrier. AGREE performs compositional analysis, allowing verification of system requirements based on composition of the component assume-guarantee contracts. By abstracting the implementation of subsystems and software components as formal contracts, large systems can be built up and verified hierarchically in the AADL model without the need to perform a monolithic analysis of the entire system at once. AGREE translates the model to the Lustre language and then performs verification using a model checker (e.g., JKind \cite{JKind}) and an SMT Solver (e.g., Z3 \cite{Z3}). 

\subsection{nuXmv}
\label{sec:nuXmv}
nuXmv is a symbolic model checker. It builds on and extends NuSMV. It implements verification for finite and infinite state synchronous transition systems. For finite-state systems, it complements NuSMV's basic verification techniques with a family of new state-of-the-art verification algorithms. For infinite-state systems, it extends the NuSMV language with new data types, namely integers and reals, and it provides advanced SMT-based model checking techniques. nuXmv implements SMT-based model checking techniques \cite{cavada2014}. 

\section{UAM Case Study}\label{sec:case-study}

Several research studies have focused on developing a formal definition for a case study \cite{runeson08}. Runeson in his research introduces case studies as a methodology and provides guidelines for the elements of a case study. Our research methodology utilizes a case study as a way to model and represent scenarios envisioned in the future for UAM. Our case study is a "Conceptual Case Study". The design of our case study is guided by the objective of contingency management for UAM to be performed by an autonomous agent. The data collection process involved requirements gathering by evaluating scenarios as described in research articles on UAM and by interacting with a human-autonomy teaming expert. Once requirements were collected, the scenarios were modeled in our framework and evidence was collected in the form of models, simulation results, and formal verification results. The collected data, the designed models, and the final results were analyzed to identify the satisfaction of results and lessons learned. Finally, reports were generated that included the models, outcomes, and the lessons learned.  

In our case study application of the AHMIIAS framework (Fig. \ref{fig:TechApproach}), we gather the requirements for human-machine interaction during the requirements phase. We use AADL with the AGREE Annex (see Sections \ref{sec:aadlintro} and \ref{sec:agree}) during the design and analysis phase to capture a formal system architecture model with requirements allocated to components. We also use the AGREE tool to perform formal verification of the architecture, showing that the system requirements are satisfied given the component requirements. The IAS agent is implemented in the cognitive architecture Soar during the implementation phase, and the IAS agent is integrated with the X-Plane environment so that we can run simulations to test the implemented behavior together with a UAM air vehicle model. Finally, to prove that the IAS requirements are satisfied by the Soar implementation, we translate the agent from Soar to the nuXmv model checker and perform formal verification over the resulting formal model in the testing and verification phase. The architecture models, IAS agent, translator code, and verified models can all be found on our project repository\footnote{https://github.com/loonwerks/AHMIIAS}.

\subsection{Example IAS Scenarios}
Scenarios were developed to determine the roles and responsibilities of a human pilot working together with an IAS to enable UAM operations. In the Unreliable Sensor Scenario, an urban canyon reduces the reliability of GPS for determining the location, leaving Lidar and IMU reliable. The IAS notices the difference between the GPS position value and the Lidar and IMU values, which indicates an unreliable GPS sensor. The IAS determines the correct position using Lidar and IMU without GPS and notifies the pilot about the unreliable GPS sensor and correct position. The pilot either a) acknowledges the unreliable GPS sensor, or b) rejects the IAS interpretation that the GPS sensor is unreliable.

In the Aborted Landing Scenario, a damaged vehicle on a landing pad prevents a safe landing. The pilot prepares for landing and notices the landing area is not suitable. The pilot calls for an aborted landing which brings up a rerouting checklist. In the rerouting checklist, the IAS reminds the pilot of the unreliable GPS and correct position. The IAS calculates routes for new landing options, presents the best option and detailed reasoning to the pilot. The pilot acknowledges the correct position and either a) accepts the new landing option and route, or b) requests alternates, sees options with IAS reasoning, and chooses an alternate landing area.

These scenarios were chosen to allow the IAS to first detect an off-nominal situation in the Unreliable Sensor Scenario, and to allow the pilot to first detect the unsuitability of the landing pad in the Aborted Landing Scenario. In both, the pilot is the Pilot In Command and the final decision-maker. The pilot can override the IAS determination of unreliable sensor, calls plays to inform the IAS of goals, and determines the safety of the landing area. The IAS monitors and assists in decision making, informs the pilot of a change in sensor reliability, and provides routes to alternate landing sites.

\subsection{IAS Implementation}

\begin{figure}[h]
\centering
\includegraphics[scale=0.30]{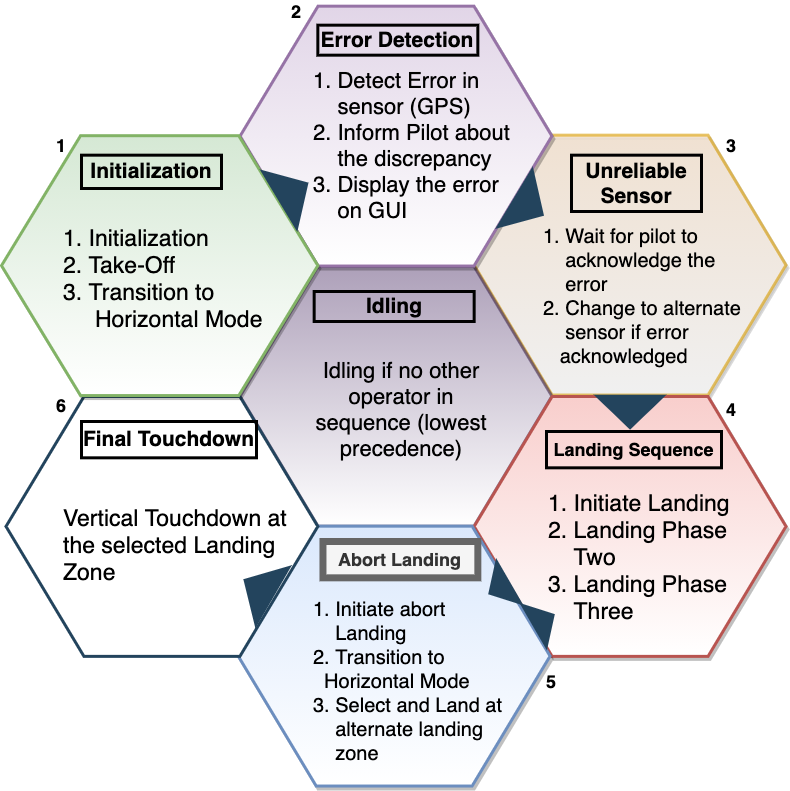}
\caption{Soar Agent Overview}
\label{fig:SAV}
\end{figure}

We implemented the IAS agent in Soar, a cognitive architecture. The cognitive model for the IAS agent consists of rules. The rules for the agent can be broken into seven categories: Initialization, Error detection, Unreliable sensor, Landing, Abort landing, Final touchdown and Idling, as shown in Figure~\ref{fig:SAV}. The rules within these categories execute actions to support the designated sequence of operations for the unreliable sensor and aborted landing scenarios, as well as takeoff and landing operations. 


\subsection{Scenario Simulations}

\begin{figure}[ht]
\centering
\includegraphics[scale=0.70]{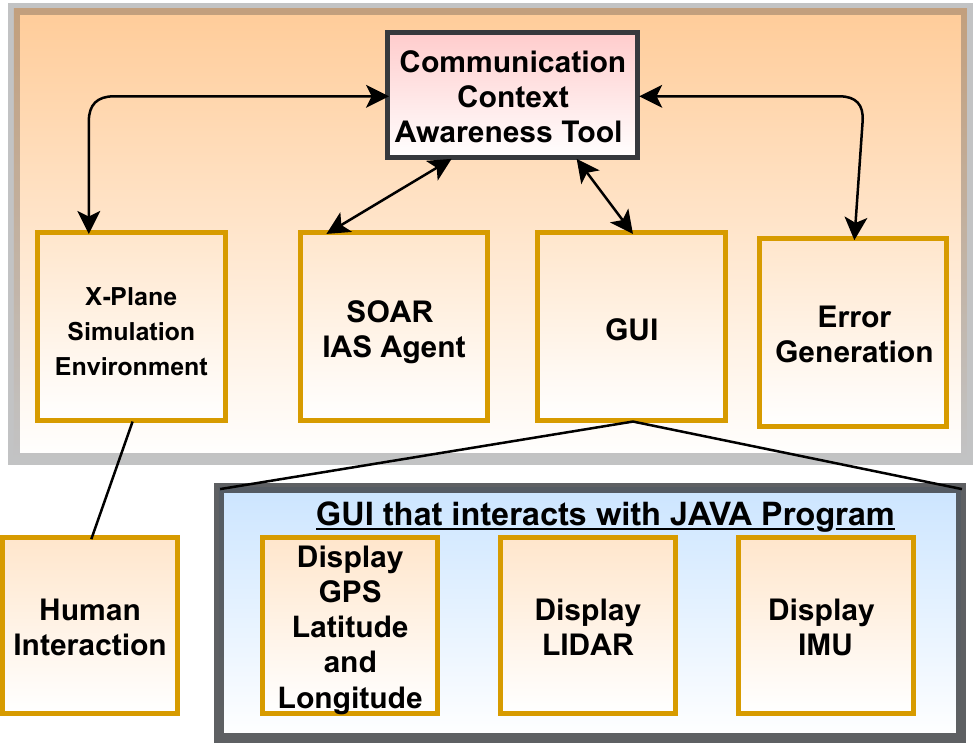}
\caption{Simulation Architecture}
\label{fig:SimArch}
\end{figure}

The simulation architecture in Figure \ref{fig:SimArch} shows the information flow between the following interacting components: the X-Plane simulation environment, the Soar IAS Agent, the Communication Context Awareness Tool (CCAT), the Graphical User Interface (GUI), and the error generation module. The X-Plane simulation environment was utilized to simulate contingency scenarios with the AW609 aircraft. Additionally, we created a GUI to interact with the aircraft. This GUI enables input from the human pilot and permits the creation of emerging situations. The Soar IAS agent implements rules that are derived from human-machine interaction research. For example, the IAS agent provides a warning to the pilot that there is a potential sensor error. The pilot provides information as to whether to abort a landing. 

\begin{figure}
\centering
\includegraphics[scale=0.18]{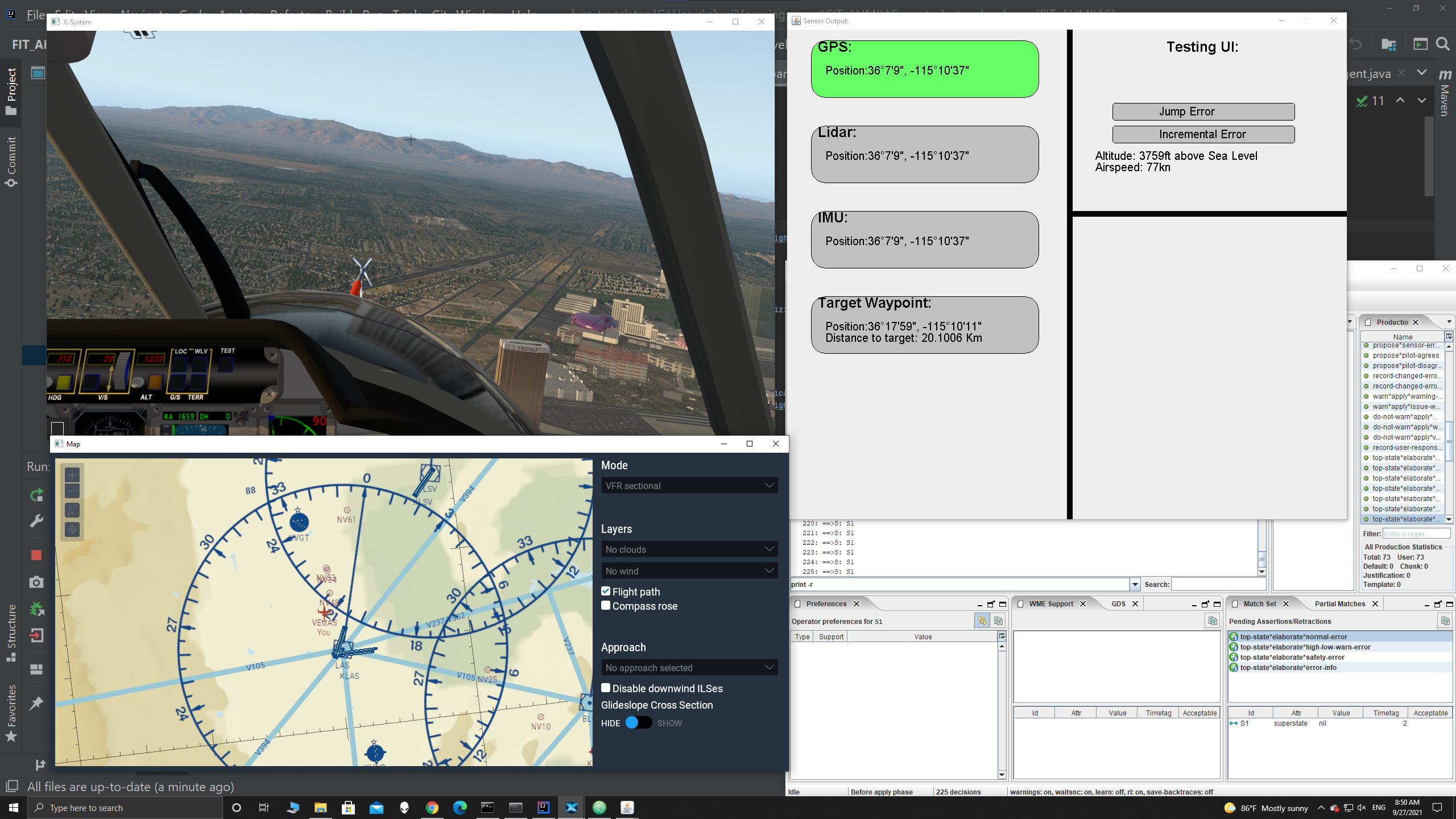}
\caption{X-Plane AW609 with GUI}
\label{fig:SimXplane}
\end{figure}

The GUI displays the values received from the GPS, Lidar, and IMU sensors (Figure \ref{fig:SimXplane}). The error generation module (Figure \ref{fig:SimXplane} ) within the Testing User Interface (UI) induces error in the value displayed from GPS. CCAT is a combination of XPC (X-Plane Connect) developed by NASA to capture information from X-Plane environment and computations performed for the unreliable sensor and abort landing scenarios. CCAT has been developed in Java programming language. The IAS agent performs actions that a human would conduct, whereas CCAT is automated technology that performs all the computations. Presently, the CCAT performs the calculations related to error among the sensors, identifying routes that traverse less populated areas, and identifying nearby airports during emerging damaged landing areas.



\subsection{Identification of Safety Requirements}

This project uses CRM to develop roles and responsibilities of an IAS assisting a human pilot acting as Pilot In Command in an UAM aircraft, and also uses CRM to generate requirements and procedures for the IAS. These requirements and procedures are used to develop the interface between the pilot and IAS. The interface is used in the simulation, formal models, and the IAS implementation. 

Here we briefly describe an example of how CRM guides interface development. Since the IAS has the ability to detect an unreliable sensor, CRM requires the IAS to present information about that sensor to the pilot and to receive the pilot’s decision on whether or not to use the sensor. The interface must allow these actions, and these actions are used to develop the formal model of the IAS. 

One example of a safety requirement is ``If the operator disagrees that the active sensor is unreliable, then the active sensor should not change." This is formalized in AGREE, and we prove using AGREE's assume-guarantee analysis that our human-machine team architectural model satisfies this requirement.

\subsection{Human-IAS Team Models}
\label{sec:AADL}
There are multiple benefits to architectural modeling and analysis:
\begin{enumerate}
    \item An architectural model helps the team agree on a common architecture and express expected interfaces unambiguously. 
    \item Using formal methods, system-level requirements (such as safety requirements) can be expressed and proven, using specified component-level requirements.
    \item A formal analysis called realizability analysis can be used to check for conflicts amongst the set of requirements for a given component.
    \item A formal model of the architecture is amenable to future automated translation to downstream component design and verification tools, helping to ensure that component requirements are properly passed down to component development teams. 
\end{enumerate}

\subsubsection{Human-Machine Team Architecture Model}
\begin{figure}[hbt!]
\centering
\includegraphics[scale=0.6]{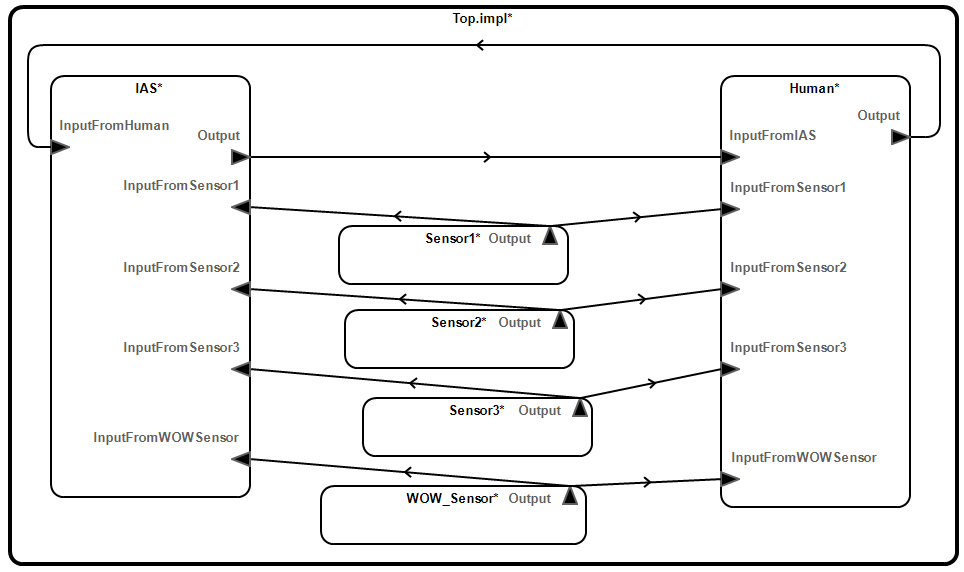}
\caption{Human-IAS Team Model}
\label{fig:HumanIASTeamModel}
\end{figure}

A key aspect of our approach is to include the human in the model. Therefore, our top-level model includes components for both the IAS and the human operator. We also include air vehicle components required for our selected scenarios. These include three position sensors as well as a Weight on Wheels (WoW) sensor/subsystem, which is used to determine when the vehicle has completed the landing phase (and an abort landing command is no longer viable). The graphical representation of the AADL model, showing the components and connections, is provided in Fig. \ref{fig:HumanIASTeamModel}. The three position sensors are labeled Sensor 1, Sensor 2, and Sensor 3 in the AADL model and represent GPS, Lidar, and IMU, respectively. The full details of the information shared between components and the current requirements on each component are captured in the textual models\footnote{https://github.com/loonwerks/AHMIIAS/tree/v1.0/architecture/AADL\_main\_project/ahmiias-architecture}. In addition to capturing the components and connections in AADL, we capture requirements for the Human, IAS, and Human-IAS Team as guarantees in the AGREE language. While we cannot place requirements on a human \emph{per se}, we can use these formalized guarantees as a means to capture expected human behavior and as a foundation for reasoning about the human-IAS team. These guarantees can be validated in a simulation environment with a human operator, and some may be enforced by the human machine interface. For example, to enforce a ``requirement" that the human operator only commands abort landing in the landing phase, a display system for the human-machine interface might gray out an abort landing option when the vehicle is not in the landing phase. Selected requirements/guarantees for the human operator and the IAS are shown in the subsections that follow.

\subsubsection{Human/Operator ``Requirements”}
A key requirement/expectation on the human operator is that he or she responds to unreliable sensor messages from the IAS. The guarantee corresponding to Sensor 1 is shown in Figure \ref{fig:HumanReqs}. Similar guarantees are in the model for Sensor 2 and Sensor 3. The situation where the operator fails to respond to an unreliable sensor message or is late in responding will be explored in future work (see Section \ref{sec:conclusion}). 


\begin{figure}[hbt!]
\centering
\begin{lstlisting}[basicstyle=\scriptsize\ttfamily]
guarantee "Respond to message that Sensor1 is unreliable":
				prev(not InputFromIAS.Sensor1_Reliable, false) <=> 
					((Output.Sensor1_Unreliable_Response = enum(Response, Agree)) 
						or (Output.Sensor1_Unreliable_Response = enum(Response, Disagree))); 
\end{lstlisting}
\caption{Example Human/Operator ``Requirement''}
\label{fig:HumanReqs}
\end{figure}

\subsubsection{IAS Requirements}
The reliability of each sensor is computed by checking whether its position output is within a horizontal and vertical threshold of at least one of the other sensor’s positions. The horizontal and vertical thresholds depend on Above Ground Level (AGL). The IAS shares its reliability assessment of each sensor with the operator.


The IAS shares both the active sensor ID and the recommended sensor ID with the operator. The active sensor stays the same unless the operator agrees that it is unreliable and there is another reliable sensor available. This is captured in the the guarantee in Figure \ref{fig:IASActiveSensor}. 

\begin{figure}[hbt!]
\centering
\begin{lstlisting}[basicstyle=\scriptsize\ttfamily]
guarantee "The active sensor should stay the same unless the operator agrees that it is 
    unreliable and there is another reliable sensor available.":
	Output.Active_Sensor = 	
		if (previous_recommended_sensor = NIL)
		    then previous_active_sensor
		else if ((previous_active_sensor = 1 and 
		            InputFromHuman.Sensor1_Unreliable_Response = enum(Response, Agree))
				    or (previous_active_sensor = 2 and 
				        InputFromHuman.Sensor2_Unreliable_Response = enum(Response, Agree))
				    or (previous_active_sensor = 3 and 
				        InputFromHuman.Sensor3_Unreliable_Response = enum(Response, Agree)))
		    then previous_recommended_sensor
		else previous_active_sensor;
\end{lstlisting}
\caption{IAS Active Sensor Determination}
\label{fig:IASActiveSensor}
\end{figure}

\subsection{Formal Verification of Human-IAS Team Requirements}
We can express several desired properties of the human-machine team such as: 
\begin{enumerate}
\item ``The operator responds to unreliable sensor alerts from the IAS."
\item ``We can't have just Sensor 1 reliable."
\item ``We can't have just Sensor 2 reliable."
\item ``We can't have just Sensor 3 reliable."
\item ``Unless the active sensor becomes unreliable, the sensor recommended by the IAS is the current active sensor."
\item ``If the operator agrees with the IAS that the active sensor is unreliable, and if the IAS recommended another sensor for use, then the new active sensor shall be the recommended sensor."
\item ``If the operator disagrees with the IAS that the active sensor is unreliable, then the active sensor should not change."
\item ``If an unreliable sensor is the active sensor, it must be the case that either the pilot disagreed with the IAS assessment or the sensor just became unreliable on this timestep or there was no reliable sensor available on the previous timestep."
\item ``The active sensor is one of the available sensors on board."
\end{enumerate}

We then formalize and prove that our human-machine architectural model satisfies these properties using AGREE's assume-guarantee reasoning. For example, the formalized statement of Property 8 in the preceding list is shown in Fig. \ref{fig:HumanIASTeamProperty}. 

\begin{figure}[hbt!]
\centering
\begin{lstlisting}[basicstyle=\scriptsize\ttfamily]
	lemma "If Sensor 1 is both the active sensor and unreliable, it must be the case that 
	    either the pilot disagreed with the IAS assessment 
		or the sensor just became unreliable on this timestep 
		or there was no reliable sensor available on the previous timestep":
				((IAS.Output.Active_Sensor = 1) and not IAS.Output.Sensor1_Reliable) 
					=>  ((Human.Output.Sensor1_Unreliable_Response = enum(Response, Disagree)) 
						    or 	prev(IAS.Output.Sensor1_Reliable, true) 
					    	or not prev(reliable_sensor_available, true));
\end{lstlisting}
\caption{Human-IAS Team Property}
\label{fig:HumanIASTeamProperty}
\end{figure}

\subsection{Soar-to-nuXmv Translator}
\label{sec:translator}
The formal analysis described in the previous section shows that IF the IAS implementation satisfies its requirements and IF the human operator satisfies his or her ``requirements," then the human-machine team will have the desired properties. One still needs to show that an IAS implementation satisfies the IAS requirements. To formally prove this, we first need to translate our IAS agent which is implemented in Soar (see Section \ref{sec:Soar}) to a formal language such as nuXmv (see Section \ref{sec:nuXmv}). We developed a Soar-to-nuXmv translator for this task. The following boxes show the IAS Agent Soar Takeoff Rule and its corresponding representation in nuXmv. 

\begin{tcbraster}[raster columns=2]
\begin{tcolorbox} [text width=6.2cm, label=rule1:A ,title = IAS Agent Soar Takeoff Rule] 
 { sp \{propose*takeoff
    \\(state $<s>$ \string^name takeoff)
    \\($<s>$ \string^flight-mode vertical)
    \\($<s>$ \string^io.input-link.flightdata $<fd>$)
    \\($<fd>$ \string^throttle $<th>$ $<$ 0.9)
\\$-->$
    \\(write (crlf) |Throttle | $<th>$)
    \\($<s>$ \string^ operator $<o>$ +)
    \\($<o>$ \string^ name takeoff)
\}}
\end{tcolorbox}
\begin{tcolorbox} [text width=8cm, title = nuXmv Model for the Takeoff Rule]
VAR soarAgent : soarRules(state\_superstate, state\_operator\_name, state\_name, state\_flight-mode, state\_io\_throttle, state\_io\_altitude, state\_io\_airspeed, state\_sensor-unreliable, state\_io\_sensor-error, state\_io\_initiate-landing, state\_landing, state\_io\_distance-to-target, state\_io\_abort-landing, state\_io\_target-altitude, state\_io\_autoflaps, state\_io\_air-brake, 
state\_io\_target-speed);


TRANS
   \\ \hspace*{0.5cm}  next (state\_operator\_name) =
    \\ \hspace*{0.7cm}case
	\\ \hspace*{1.2 cm}(state = run \& state\_name=takeoff \& 
	\\ \hspace*{1.2 cm}state\_flight-mode=vertical \& 
	\\ \hspace*{1.2 cm}state\_io\_throttle$<$0.9): takeoff;
	\\ \hspace*{1.2 cm}TRUE : state\_operator\_name;
    \\ \hspace*{0.7 cm} esac;
\end{tcolorbox}
\end{tcbraster}

In the translation process, the first step involves identifying all the variables (operators, input/output data) and expanding the shorthand notations that Soar uses. For example, $<s>$ is a representation of the present state, which is expanded to state; and $<o>$ is the shorthand for an operator, which is replaced with operator during the translation process. The left-hand side of the $\rightarrow$ consists of the condition that needs to be true for the right-hand side to be executed. In the Soar rule for Takeoff, the left-hand side indicates the state name should be Takeoff, the flight mode should be vertical, and input flight data representing throttle should be less than 0.9 to execute the right-hand side of the rule, which changes the state operator name to takeoff. In nuXmv the IAS agent has two states: \textit{Start} and \textit{Run}. In the \textit{Start} state, the conditions for execution of all the operators are evaluated to check which one to select for execution, which is similar to how Soar operates. In the \textit{Run} state, the selected rule is applied. The conditions of a Soar rule are translated into conditions of a case statement within a transition statement in nuXmv. The updated values in nuXmv are based on value changes made in the actions of the Soar rule.

The algorithm for the translator is shown in Algorithm 1. We define a Soar production rule as a function of a finite set of variables $v_i$ $\in$ V, where i = 1, 2, 3, ... n, whose valuation val(V) = $v_i$  represent the state of the system along with a finite set of well-formed formulae (WFF) $\phi = \{\phi_1,\phi_2,...\phi_m\}$, representing the left-hand side of the Soar production rule (e.g., the preconditions), and a finite set of WFF $\psi= \{\psi_1,\psi_2,...\psi_r\}$, representing the actions embodied by the right-hand side of the Soar production rule. 
The input includes the rules from the Soar model represented as a tuple, $rname(V,(pre\{\phi_1, \phi_2, ... \phi_{m}\},$ $post\{\psi_1, \psi_2, ... \psi_{r}\}))$ 
These Soar rules are translated into Infinite State Machines $ISM = (S, S_0, Vars, G, $ $Act, Tr)$, where S is the set of states, $S_0$ is the initial state, Vars represent the variables, and G represents the guard conditions. It is assumed that the preconditions and postconditions within the Soar rules are well-formed formulas.

Steps 1-14 involve identifying, declaring, and, for symbolic constants, listing the values for all the variables that exist within the Soar rules. Steps 15-18 involve initializing the ISM with its states, variables, guard conditions, transitions, and actions. Steps 19-24 include generating the MODULE that controls the cycle of selecting one of the proposed rules and then applying the rule, as is done in Soar. During the selection process, the ISM transitions from the $start$ state to the $run$ state based on the satisfaction of a precondition $pre(\phi_i)$. Then the selected rule is applied, when at the run state, based on the satisfaction of the postconditions $post(\psi_i)$, which are represented as guards. 

Steps 25-33 involve the generation of the state operator that needs to be executed. While generating the state operator name, the satisfaction of the precondition is checked along with any priorities associated with the value of the state operator. If the state operator has an associated priority, it is generated at the top of the list; otherwise, it is generated at the bottom. Presently, the algorithm only performs binary priorities, i.e., with or without priority.
The change in the values of all the other variables is performed within Steps 34-39 based on the evaluation of the postcondition.

\begin{algorithm}[!]
\small
 \caption{Generate Infinite State Machine $ISM = (S, S_0, Vars, G, Act, Tr)$ from $rname(V,(pre\{\phi_1, \phi_2, ... \phi_{m}\}, post\{\psi_1, \psi_2, ... \psi_{r}\}))$}
 \label{alg1}
 \begin{algorithmic} [1]
 \FORALL{$i \in \{1, \ldots, m \}, j \in \{1, \ldots, r \}$ }
 \FORALL{$Vars\in pre\{\phi_i\}, post\{\psi_j\}$}
 \STATE $EXTRACT Vars \gets \{var_1, var_2, \ldots, var_i, \ldots, var_n\}$:
 $where var_i: type \{integer, real, symbolic\ constant\}$
    \IF{$\{var_i: type == symbolic\ constant\}$} 
    \STATE assign value $var_i \gets \{v_i\}$;
    \IF{$\{var_i == var_j\}$} 
        \STATE assign value $v_j$ to list $var_i \gets \{v_i...v_j \}$; 
        \STATE remove $var_j$;
        \ENDIF
        \ELSIF {$\{var_i == var_j\}$}
        \STATE remove $var_j$;
       \ENDIF
 \ENDFOR
 \ENDFOR
 \FORALL{Vars}
\STATE ASSIGN 
$INIT var_1 \gets v_1, INIT var_2 \gets v_2, $
\ldots
$INIT var_n \gets v_n$\\
\ENDFOR

 \STATE Initialize $ISM_i$ = ( S $\gets \{Start,Run\}$, $s_0 \gets \{Start\}$, $Vars =\{state\_superstate = nil\}$ G = $\{\}$, Act = $\{\}$, Tr = $\emptyset$)
 
 VAR SoarModuleInstance\{Vars\}
 \STATE MODULE: VAR state: \{start, run\};
\STATE next (state) = 
        Case
            \FORALL{$i \in \{1, \ldots, m \}, pre\{\phi_i\}, post\{\phi_i\}$}
                \STATE $Tr \gets (state = start \ \& \ G_i == pre(\phi_i) : run)$;
                 \STATE $Tr \gets (state = start \ \& \ G_i == post(\psi_i) : run)$;
            \ENDFOR
 \FORALL{$i \in \{1, \ldots, m \}, state\_operator\_name \in pre\{\phi_i\}$}
                \IF{\{state\_operator\_name count == 1\}} 
                \STATE Print TRANS
                \ENDIF
                    \STATE $Tr \gets (next (state\_operator\_name)= 
                    \{state == {run} \ \& \ G_i == {pre(\phi_i)}\} : v_i)$
                    
                    \IF{$v_i \ has \ priority$} 
                    \STATE Generate the $next (state\_operator\_name)$ statement on top;
                    \ELSE
                   \STATE Generate the $next   (state\_operator\_name)$ statement on the bottom;
                    \ENDIF
            \ENDFOR
            
\FORALL{$i \in \{1, \ldots, r \}, Vars \in post\{\phi_i\}$}
                \IF{\{$var_i$ count == 1\}} 
                \STATE Print TRANS
                \ENDIF
                \STATE $Tr \gets (next (var_i)=  state = run \ \& \ G_i = (\psi_i) : v_i)$
                \STATE $Act \gets Act \ \land \ (next(var_i) = v_i)$
            \ENDFOR
\end{algorithmic}
\end{algorithm}

\subsection{Formal Verification of IAS Requirements}

The high-level requirements for the IAS captured in AGREE must be verified on the IAS implementation. This is an important part of a complete assurance argument for the human-IAS team. For example, we map the requirement for IAS Active Sensor Determination (Fig. \ref{fig:IASActiveSensor}) to the following nuXmv property:
\begin{verbatim}LTLSPEC (state_io_sensor-to-use = nil U (state_io_pilot_decision = agree 
& (state_operator_name = gps-sensor-error-over-limit 
   | state_operator_name = lidar-sensor-error-over-limit 
   | state_operator_name = imu-sensor-error-over-limit)))
\end{verbatim}
The nuXmv property checks that the IAS does not change the sensor to use until a sensor is faulty and the pilot agrees. It has been verified with nuXmv over our translated Soar agent. This property captures the high-level intent of the corresponding AGREE requirement but is not a perfect semantic match. We discuss this further in lessons learned. 

The formal verification of the IAS agent in nuXmv used an input template that represents the dynamics of the AW609 as obtained from X-Plane. For example, we included the relationship between the throttle and the altitude, as well as the threshold values that indicate error in sensors. The verification of the IAS agent was also performed without the input template; this resulted in generation of counterexamples such as 1) the altitude remained at zero even though the throttle reached its highest value and 2) the sensors were in error but returned to normal before diagnosis.
The categories of verification performed were: reachability, checking invariants, checking normal execution, and responding to off nominal situations. Examples are provided below. The total number of queries executed for the unreliable sensor scenario is 24 and that for the abort landing scenario is 26. The maximum number of steps of execution for a query for the unreliable sensor scenario is 67 and the minimum is 6. The maximum number of steps of execution for a query for the abort landing scenario is 140 and the minimum is 44.
\begin{itemize}
    \item \textit{Reachability}: \textbf{LTLSPEC F q, in future q holds}, where q can be (state\_io \_altitude $>$ 10000) or (state\_operator\_name = transition)
    \item \textit{Invariants}: \textbf{INVARSPEC q, invariant q }is satisfied, where q can be (state\_io\_throttle $<=$ 1.0)
    \item \textit{Handling off nominal situations}: \textbf{LTLSPEC F(X p$->$q (next p leads to q)) or (p U q) (p until q) or (p S q) (p since q)}, where p$->$q can be of the form, counter\_detect\_transition\_to\_lidar $<=$ 5 $->$ state\_io\_sensor-to-use = lidar)
    \item \textit{Handling normal operations}: \textbf{LTLSPEC F(X p$->$q) or (p U q) or (p S q) }, where p$->$q can be of the form, state\_operator\_name = state\_io\_throttle $<$ 1.00  $->$ state\_flight-mode = horizontal)
    \item \textit{Response to occurrence of event} \textbf{LTLSPEC G(p $->$F q), Globally p leads to q in future}, where p is (state\_operator\_name = gps-sensor-error-over-limit) and q is (state\_sensor-unreliable = yes)
\end{itemize}

\subsection{Results and Lessons Learned} \label{sec:results}
The iterative process of formal verification and simulation helped identify flaws in the design of the IAS agent and the human-IAS interactions. Table \ref{table:findings} shows the findings.

\begin{table}[h]
\centering
\begin{tabular}{|p{0.7cm}|p{2.0cm}|p{9cm}|} 
 \hline
No. & Error Type & Findings\\
 \hline
1. & IAS Design Error & Soar agent missed response to human agree - disagree rule, as it was being handled at the CCAT interface.\\
\hline
2. & IAS Design Error & state\_operator\_name throttle case condition missed the equal logical operator(th $<$ 0.9 \&  th $>$ 0.9). As a result, the throttle value was exceeding 1.0, which is an error.\\ \hline
3. & IAS Design Error & SOAR agent missed the human selection response after abort landing. \\ \hline
4. & Translation Error & Superstate, a state in Soar before Soar graph is generated only indicates that the Soar graph exits or not, it does not need to be translated, but was translated \\ \hline
5. & IAS Design Error & Soar agent missed a case statement to set abort\_landing to ``NO" after it has been addressed.  \\ \hline
6. & Translation Error & Type of some of the variables were generated as integer, but were used as real, it was detected through properties that proved immediately\\
\hline
7. & IAS Requirements Error & Parentheses/order of operations error with regard to selecting the recommended sensor\\ \hline
8. & Operator ``Requirements" Error & Selection of landing option was occurring too late, one time step after the options were ready.\\ \hline
9. & IAS Design Error & IAS does not check that the sensor it is switching to is reliable before recommending a switch to the pilot. This is a divergence from the IAS requirements as specified in AGREE. \\ \hline
\end{tabular}
\caption{Findings after architecture and formal verification}
\label{table:findings}
\end{table}

The lessons learned from the application of our methodology to the UAM case study are:
\begin{enumerate}
\item AGREE-to-nuXmv mapping of IAS requirements: Several of the IAS requirements as captured in AGREE do not have a direct semantic mapping to nuXmv. This is due in part to the fact that the initial IAS agent was designed in parallel with the human-IAS team model in AADL/AGREE. Hence the interface and expected IAS agent behavior, as captured in AGREE, are not the same as those of the IAS implementation in Soar. For example, the IAS does not have a notion of a recommended sensor. Another challenge is that the formalisms of AGREE and nuXmv are not the same. AGREE uses Past-time Linear Temporal Logic (PLTL) whereas nuXmv uses Linear Temporal Logic (LTL) and Computational Tree Logic (CTL). While PLTL and LTL have the same expressiveness, it is not straightforward to refer to the prior value of a variable in a nuXmv property. Nonetheless, we can map the intent of each AGREE requirement to its closest analog in nuXmv and check its validity. 
\begin{itemize}
    \item Solution: Future work includes better aligning the architectural model and the implementation. An ideal workflow would build the IAS implementation using the interface and requirements first specified in the architectural model, and these would be kept in sync as refinements occur.
\end{itemize}
    \item Interaction delay: There is a potential of having a delay between the communication that can occur between the IAS agent, CCAT and the pilot. This should be considered while designing the interactive system. For example, we identified that the IAS agent would repeat the execution of operations due to a delay between Soar issuing a command to Java and the command taking effect.
\begin{itemize}
    \item A solution: A copy of the output command was stored and used to prevent repeat operations until the command took effect.
\end{itemize}

\item Variable type declaration: Variables are not typed in Soar. So, assignment of values to variables need to be evaluated during a full pass through the Soar model to identify the type. For example: An incorrect declaration of a variable (e.g., state\_io-air-brake) caused all properties that should not prove to prove. 
\begin{itemize}
    \item Solution: Found and rectified during property verification.
\end{itemize}

\item Separation of responsibilities: Proper allocation of tasks needs to be completed before implementation. Since the IAS agent is expected to perform tasks performed by the human, we need to carefully identify all the computation related tasks and create automation to handle computation separately. Otherwise, the heterogeneous mix of tasks leads to a challenging situation for verification. For example, comparison of error differences among the sensors was earlier performed by Java XPC, which was an inefficient design according to the principle of separation of responsibilities. This was captured during property verification.
\begin{itemize}
\item Solution: Rectified with proper allocation of tasks. SOAR agent performs the error check, whereas the calculations are done in JAVA-XPC.
\end{itemize}
\item Mapping from architecture to implementation: One to one algorithmic mapping from AADL/AGREE to nuXmv needs to be developed as that will capture errors in design when transitioning from architecture to agent implementation and formal verification. For example, before changing from an erroneous sensor in AADL/AGREE, reliability of the new sensor is checked, but this second check is not performed in the IAS agent. This was captured when mapping from AADL/AGREE to nuXmv was performed. 
\begin{itemize}
\item Solution: Algorithmic mapping from AADL/AGREE to IAS model.
\end{itemize}
\end{enumerate}

\section{Conclusions and Future Work}
\label{sec:conclusion}
Our AHMIIAS assurance framework, which integrates human-machine interactions in a formal model, helped identify and validate the responsibilities of the IAS and the human. The responsibilities for the IAS mostly focused on maintaining situational awareness, taking actions in normal situations, and taking actions under contingency if commanded by the pilot. The responsibilities for the IAS were identified through iterative discussions with a human-autonomy teaming expert, which resulted in following the philosophy that the human always has the final authority.  We demonstrated how human-IAS interactions can be modeled early in the design phase for architectural analysis. Then, the requirements were validated through detailed implementation of algorithms in the simulation environment that integrated X-Plane with the IAS agent implemented in Soar.  Finally, our approach illustrated transitioning from  simulation to formal verification through automated translation of the IAS agent from a cognitive model to a formal verification environment. We identified several errors by using this approach and we captured several lessons learned. 

Our future work will explore how our assurance framework can be extended to accommodate learning mechanisms. We will identify the human machine interactions that need to be implemented for a learning system and extend our translation algorithm and verification approach to accommodate an IAS agent that learns. Another area of future work is to explore what happens when the human operator or IAS violates one of its requirements/expected behaviors. For example, the IAS may have a subcomponent hardware failure or the human may have a high workload and not respond to an alert from the IAS. We plan to leverage the Architectural Modeling and Analysis for Safety Engineering (AMASE) tool \cite{StewartAMASE} to reason about the human-machine team properties in the presence of faults.  

\bibliographystyle{eptcs}
\bibliography{generic}
\end{document}